# First-principles study of phononic thermal transport in monolayer $C_3N$: a comparison with graphene

Yan Gao[a], Haifeng Wang,[a]* Maozhu Sun[a], Yingchun Ding[b], Lichun Zhang[c], Qingfang Li[d] and Jidong Zhang[a]

[a]*Department of Physics, College of Science, Shihezi University, Xinjiang 832003, China. Email: whfeng@shzu.edu.cn*

[b]*College of Optoelectronics Technology, Chengdu University of Information Technology, Chengdu, 610225, China*

[c]*School of Physics and Optoelectronic Engineering, Ludong University, Yantai, 264025, China*

[d]*Department of Physics, Nanjing University of Information Science&Technology, Nanjing 210044, China*

Very recently, a new graphene-like crystalline, hole-free, 2D-single-layer carbon nitride $C_3N$, has been fabricated by polymerization of 2,3-diaminophenazine and used to fabricate a field-effect transistor device with an on-off current ratio reaching $5.5 \times 10^{10}$ (Adv. Mater. 2017, 1605625). Heat dissipation plays a vital role in its practical applications, and therefore the thermal transport properties need to be explored urgently. In this paper, we perform first-principles calculations combined with phonon Boltzmann transport equation to investigate the phononic thermal transport properties of monolayer $C_3N$, and meanwhile, a comparison with graphene is given. Our calculated intrinsic lattice thermal conductivity of $C_3N$ is 380 W/mK at room temperature, which is one order of magnitude lower than that of graphene (3550 W/mK at 300 K), but is greatly higher than many other typical 2D materials. The underlying mechanisms governing the thermal transport were thoroughly discussed and compared to graphene, including group velocities, phonon relax time, the contribution from phonon branches, phonon anharmonicity and size effect. The fundamental physics understood from this study may shed light on further studies of the newly fabricated 2D crystalline $C_3N$ sheets.

# 1. Introduction

In 2004, graphene is firstly exfoliated from graphite,[1] from that moment, graphene material has attracted increasing attention because of its excellent electrical, mechanical and thermal properties.[2-5] However, graphene present zero-band-gap semiconducting electronic character, which limit the suitability for its further applicationsis. Consequently, great efforts have been made to modify graphene and further broaden its application recently.[6-8]

Nitrogen (N)-doped graphene (N-substituted or nitrogenated graphene) has become a new class of graphene material with semiconducting electronic properties, well suited for a wide range of applications.[9-12] During the past two decades, numerous carbon nitride films with different N/C ratios were widely studied, but few of them are two-dimensional (2D) crystalline layered materials.[9,13] In theory, first-principle simulations have investigated the structure and stability of many ordered 2D carbon nitride structures, while few of them can be realized in experiment at present.[14,15] Graphitic carbon nitride (g-$C_3N_4$) is a crystalline material that has been synthesized for a long-period by polymerization of cyanamide, dicyandiamide or melamine,[16] which is a semiconductor with a direct bandgap of 2.76 eV and has shown great potential in many energy applications, such as photocatalysis,[17,18,19] hydrogen generation[20] and energy storage.[21] In 2015, another 2D crystalline and ordered carbon nitride material termed $C_2N$-$h$2D was successfully synthesized via a simple wet-chemical reaction.[22] It contains a large number of regular holes in the crystalline structure, and is also conformed as a semiconductor with direct bandgap of 1.96 eV. A field-effect transistor (FET) made of $C_2N$-$h$2D exhibits an on/off current ratio of $10^7$.[22]

Very recently, a new crystalline, hole-free, 2D-single-layer carbon nitride $C_3N$ has been fabricated by polymerization of 2,3-diaminophenazine and used to fabricate a field-effect transistor device with an on-off current ratio reaching $5.5 \times 10^{10}$.[23] Successful experimental synthesis of 2D graphene-like crystalline $C_3N$ sheets consequently raise the importance of the evaluation of its intrinsic properties. In one recent theoretical study, H. Wang et al. predicted this newly synthesized 2D material owns high stiffness, superior stability and bending Possion's effect.[24] Thermal transport property is a significant factor for the application of materials. Performance of electronic devices strongly depends on high thermal conductivity for efficient heat dissipation, while low thermal conductivity is preferred in thermoelectric applications. Actually, the thermal conductivity of the other two synthesized carbon nitrides g-$C_3N_4$ and $C_2N$-$h$2D were well studied by many works.[25-29] Considering the superior mechanical and thermodynamical properties of monolayer $C_3N$, especially the higher stiffness and stability than graphene,[24] it is natural to raise the question of whether $C_3N$ could also perform well as a thermal conductor, and what level the thermal conductivity of $C_3N$ could approach. Therefore, comprehensive understanding on the thermal transport of the newly fabricated semiconductor $C_3N$ is very essential, and a comparison with graphene is helpful in revealing the underlying mechanism.

## 2. Computational details

The first-principles calculations are performed by using the density functional theory as implemented in the Vienna ab-initio simulation package (VASP)[30-31] with a plane wave energy of up to 500 eV in the expansion of the electronic wave function. The Perdew-Burke-Ernzerhof (PBE) of generalized gradient approximation[32] is chosen as the exchange-correlation functional. The internal coordinates and lattice constants are optimized until the Hellman-Feynman forces acting on each atom became less than $10^{-3}$ eV Å$^{-1}$. The convergence for energy is chosen as $10^{-6}$ eV between two steps, and a Monkhorst-Pack $k$-mesh of 13×13×1 is used to sample the Brillouin zone in the structure optimization.

An iterative self-consistent method is used for solving the phonon Boltzmann transport equation to calculate the lattice thermal conductivity with the ShengBTE code.[33] It is based on the second-order (harmonic) and third-order (anharmonic) interatomic force constants (IFCs) combined with an iterative self-consistent algorithm to solve the Boltzmann transport equation, and has also successfully predicted the thermal conductivities of many materials.[28,34-40] Phonon frequencies and the harmonic IFCs are obtained by density functional perturbation theory (DFPT) using the PHONOPY program,[41] and the 4×4×1 supercells with 5×5×1 q-meshs are used for both monolayer $C_3N$ and graphene. The third-order anharmonic IFCs are calculated by using a supercell-based, finite-difference method,[42] and the same 4×4×1 supercells with 5×5×1 $q$-meshs are used. We include the interactions with the fifth nearest-neighbor atoms for both $C_3N$ and graphene. The convergence of thermal conductivity with respect to $q$-points is tested in our calculation. The same discretization of the Brillouin zone (BZ) into a Γ-centered regular grid of 110×110×1 q-points are introduced for both $C_3N$ and graphene. The nominal layer thicknesses $h$=3.20, 3.35 Å for $C_3N$[24] and graphene[40] are used in our calculation.

## 3. Results and discuss

### 3.1 Structure and phonon dispersion

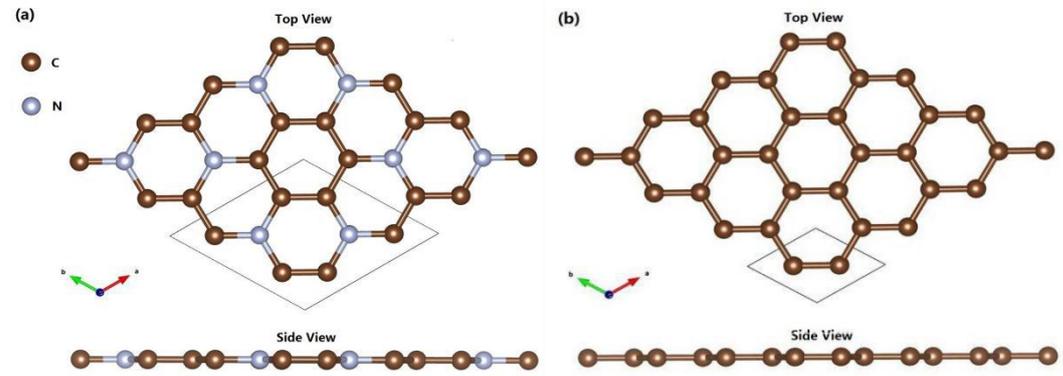

Fig. 1 Top and side view of monolayer (a) $C_3N$ and (b) graphene, the corresponding primitive cells are marked with solid lines.

The atomic structures of monolayer $C_3N$ and graphene are illustrated in Fig.1. Both of them possess the same P6/mmm symmetry (space group ID 191) with planar hexagonal structure. In fact, $C_3N$ can be considered as a nitrogen doped graphene where the nitrogen atoms substitute the native carbon atoms in the pristine graphene. In the structure of C₃N, the C-C and C-N bond are almost equivalent lengths, i.e., 1.404 and 1.403 Å. Compared to C-C bond length in graphene (1.425 Å), the bond lengths of C₃N are slightly shorter. The unit cell of C₃N contains 8 atoms, as denoted by solid box in Fig. 1(a) in which the C to N ratio is 3:1.

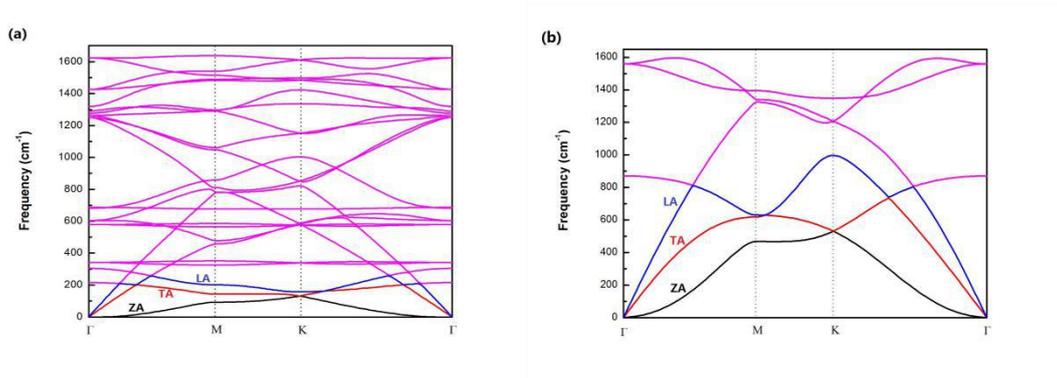

Fig. 2 The phonon spectra of (a) C₃N and (b) graphene along high symmetry directions.

The phonon dispersion determines the allowed three-phonon scattering processes and plays a significant role in precise calculation of phonon transport properties. In Fig. 2, we plot the phonon dispersions of C₃N and graphene by solving the eigenvalues of the harmonic IFCs, which are in good agreement with previous works.[24,40] Firstly, one can be see that the phonon dispersion of C₃N has no imaginary frequencies, indicating the dynamically stable of the newly fabricated 2D structure. Similar to graphene, the optical phonon branches of C₃N also has quite high eigenvalues (the highest phonon frequency in C₃N is 1638 $cm^{-1}$ and that of graphene is 1598 $cm^{-1}$). This indicates monolayer C₃N is thermodynamically stable and the bondings in C₃N are even stronger than that in graphene. Actually, according to the study of H. Wang et al.,[24] C₃N can withstand high temperature up to 2000 K and its Young's modulus (1090.0 GPa) is even higher than that of graphene (1057.7 GPa).

Each primitive cell of monolayer C₃N consists of 8 atoms (6 carbon atoms and 2 nitrogen atoms), and thus possesses 3 acoustic and 21 optical phonon (OP) branches. An important feature of the phonon spectrum in C₃N is that both the longitudinal acoustic (LA) and transverse acoustic (TA) branches are linear near the Γ point, while the z-direction acoustic (ZA) branch is quadratic near the Γ point, which is similar to graphene and other typical 2D materials, such as h-BN and MoS$_2$.

## 3.2 Lattice thermal conductivity

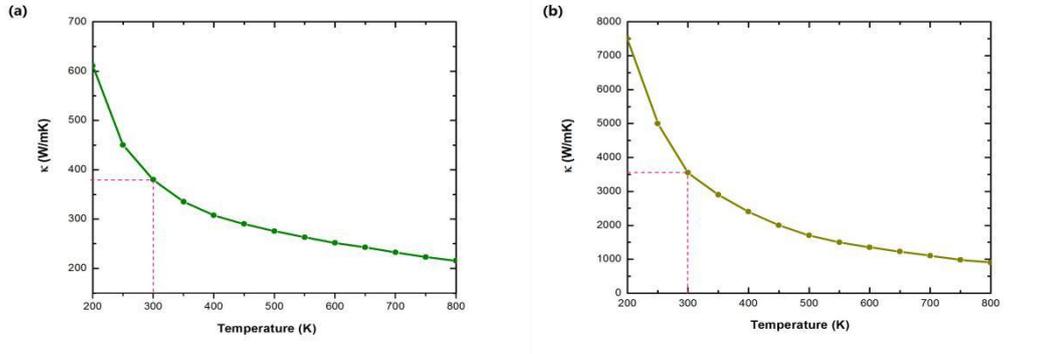

Fig. 3 Thermal conductivity as a function of temperature for (a) C$_3$N and (b) graphene. Note the different scales along the vertical axis.

From the harmonic and anharmonic IFCs, the lattice thermal conductivities of monolayer C$_3$N and graphene are obtained, as ploted in Fig. 3 (a-b). Both of their thermal conductivities decrease with an increase in temperature following the T$^{-1}$ relationship, implying the thermal conductivities are given by the anharmonic phonon-phonon interactions. At room temperature, the obtained lattice thermal conductivity of C$_3$N is 380 W/mK, and that of graphene is 3550 W/mK. Our calculated thermal conductivity of graphene agrees well with other theoretical and experimental work.[43-45] The thermal conductivity of C$_3$N is about one order of magnitude lower than that of graphene, but compared with many other typical 2D materials the value is relatively large. For example, the thermal conductivity of monolayer *h*-BN is calculated about 250 W/mK at room temperature using the same iterative self-consistent method,[36] and that of monolayer MoS$_2$ is about 100 W/mK at 300 K.[37] Especially, the thermal conductivity of C$_3$N is much larger than other 2D carbon nitrides. It's reported that the thermal conductivity of g-C$_3$N$_4$ was predicted to be 3.5-7.6 W/mK at 300 K by using the non-equilibrium molecular dynamics (NEMD) method.[25] The thermal transport properties of C$_2$N-*h*2D was calculated as 40-64.8 W/mK at 300 K by using NEMD method[26,27] and 82.22 W/mK by iterative

self-consistent method.[28]

Thus it's certain that in response to the thermal management concerns in various applications, particularly in nanoelectronics, semiconducting and high thermal conductivity of $C_3N$ nanofilms may perform as well as or even better than graphene. Because graphene nanofilms may cause undesirable effects in electronic devices owing to its high electrical conductivity, while $C_3N$ nanofilms can offer high thermal conductivity and at the same time remain electrically insulators.

**3.3. Mode level analysis**

In order to further explore the physical insight of the heat transport of $C_3N$, we perform a detailed mode level analysis. Firstly, we investigate the branch decomposed thermal conductivity. In Table 1, the contributions of different phonon modes to the thermal conductivity of $C_3N$ and graphene at room temperature are illustrated. It can be seen that the acoustic phonon modes dominate the thermal conductivity of graphene, especially, the ZA phonon modes contribute 80.1% to the total thermal conductivity. Correspondingly, the contribution of optical phonon modes is negligible (1.1%), which is also reported by many other reports[40,41] and to a certain extent, the small contribution of optical modes is a typical feature of most 2D nanomaterials, such as graphene,[40,43] $MoS_2$,[37] silicene[40,46] and black phosphorus.[47] Generally speaking, this phenomenon is mainly induced by the relatively low group velocities and short mean free path of optical phonon branches. While for $C_3N$, both the ZA and OP modes play a crucial role on the thermal transport, while the contributions from the TA and LA are quite small (3.3%). This finding is similar with the previous theoretical prediction of the phonon mode property of $C_2N$-$h$2D,[28] where the ZA and optical parts of the phonon modes together contribute to about 92.8% of the total thermal conductivity.

Table 1 Contributions from phonon branches to the total thermal conductivity of $C_3N$ and graphene (300 K).

|           | ZA    | TA    | LA   | Optical |
|-----------|-------|-------|------|---------|
| C$_3$N    | 63.2% | 2.1%  | 1.2% | 33.5%   |
| Graphene  | 80.1% | 12.1% | 6.7% | 1.1%    |

To find the reasons for understanding the significant contribution from optical branches to the total thermal conductivity on C$_3$N, we plot the frequency-dependent group velocities ($v$) and phonon relaxation time ($\tau$) of each phonon branch for the two materials in Fig. 4 according to the definition of thermal conductivity $\kappa = \sum_i c v_i^2 \tau_i$.

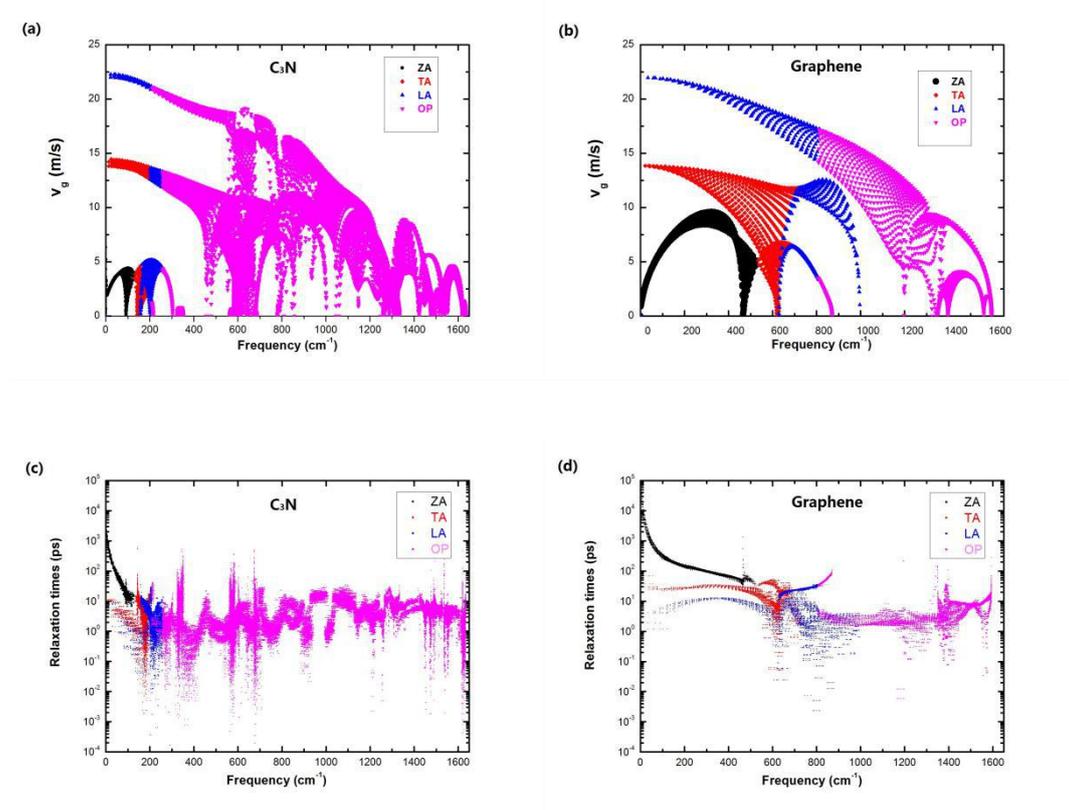

Fig. 4 (a-b) The group velocity and (c-d) the phonon relaxation time of different phonon branches as a function of frequency (300 K) for C$_3$N and graphene.

From Fig. 4(a-b), one can see that the group velocities of C$_3$N are on the same order of magnitude as those of graphene. For example, at the long-wavelength limit, the group velocities of TA and LA of C$_3$N are about 14.3 and 22.2 km/s, respectively, while they are about 13.8 and 22.0 km/s for graphene. On the other hand, great difference between $C_3N$ and graphene exists in the contribution of each phonon branch to the total group velocities, the group velocity of the ZA mode of graphene is obviously larger than that of $C_3N$. On the contrary, the group velocities of optical modes of $C_3N$ are greatly higher than that of graphene, meanwhile, the frequency ranges of optical modes are much wider than that of graphene (see Fig. 2). Undoubtedly, all of these reasons lead to the larger contribution of optical modes to the total thermal conductivity of $C_3N$.

The phonon relaxation times of different phonon branches for $C_3N$ and graphene are illustrated in Fig. 4(c-d), it can be clearly seen that the relaxation times of the out-of-plane ZA modes have a relatively longer lifetime than the other phonon modes for both graphene and $C_3N$, which is attributed to the perfect plane structure and geometric symmetry. And this just the reason of the ZA mode of $C_3N$ also makes a dominant contribution (63.2%) to its total thermal conductivity. Furthermore, compared with graphene, the phonon relaxation times of the three acoustic phonon modes of $C_3N$ are substantially reduced, which is mainly caused by the strong hybridization among the optical phonon branches. Thus we can conclude that the low group velocity, combined with the enhancement of phonon-phonon scattering, leads to a reduction of the lattice thermal conductivity of $C_3N$ compared to graphene.

**3.4 Phonon scattering process**

It's well know that the phonon-phonon scattering process depends on two factors:[40,48,49] (i) phase space for three-phonon processes, i.e., the number of channels available for a phonon to get scattered, which is determined by whether three phonon groups exist that can satisfy both energy and quasimomentum conservations, and (ii) the anharmonicity of phonon modes, i.e., the strength of each scattering channel, which is described by the mode Grüneisen parameter. To further explore the phonon-phonon scattering properties of $C_3N$ and compared to graphene, it's worthwhile to give a thorough discussion of the phase spaces and mode

Grüneisen parameters for both materials in the following section.

### 3.4.1 Phase space

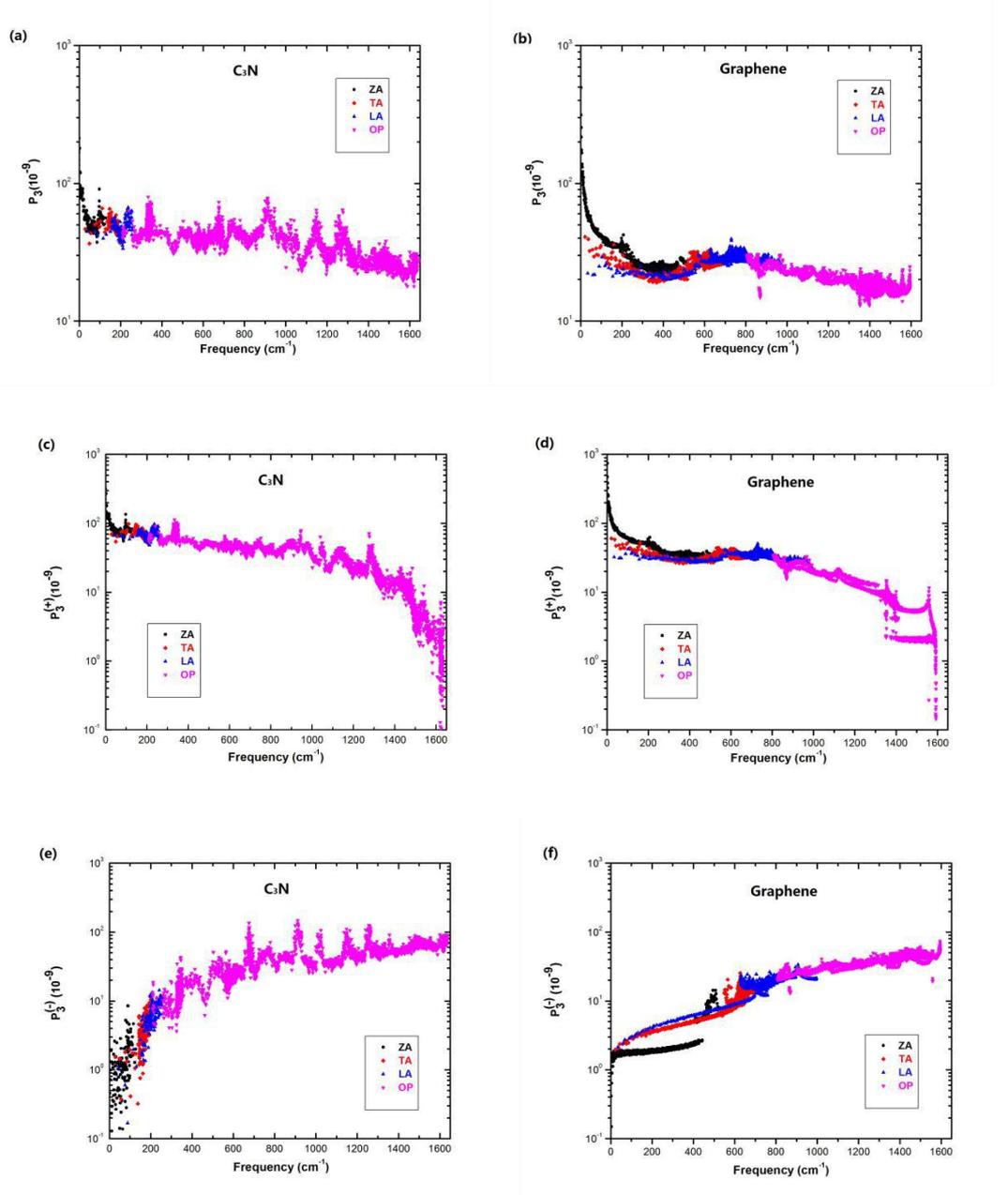

Fig. 5 Frequency-dependence of three-phonon-scattering phase space at 300 K for (a-b) total three-phonon processes (c-d) absorption processes, and (e-f) emission processes of $C_3N$ and graphene, respectively.

The total three-phonon-scattering processes ( $P_3$ ) show all available three-phonon interacting channels in the heat transfer, which is consists of two independent scattering channels, i.e., the adsorption process ($P_3^{(+)}$) and emission process ($P_3^{(-)}$). A less restricted phase space results in a large scattering rate, eventually leading to small thermal conductivity. As shown in Fig. 5(a-b), the total phase space of $C_3N$ is slightly larger than graphene, especially for the optical phonon branches, which is consistent with the fact that the thermal conductivity of $C_3N$ is much lower than that of graphene. Another fact that can be seen from Fig. 5(a-b) is that the acoustic phonon branches in graphene own obviously larger scattering phase space compared to optical phonon branches, while for $C_3N$, the optical phonon branches also possess large scattering-phase space, especially for the low frequency optical phonon branches.

For the adsorption process of three-phonon-scattering phase space ($P_3^{(+)}$), the ZA, TA and LA phonons of graphene are the main absorption channels, as shown in Fig. 5(d), i.e., the scattering channel of A+A→A is the primary three-phonon absorption process in the thermal transport of graphene.[40,43] While in the case of $C_3N$, the optical phonon branches of low frequencies also contribute greatly to the absorption processes [Fig. 5(c)], they are mainly involved in the three-phonon interactions like TA/LA/ZA + O→O processes.

The three-phonon emission channels ($P_3^{(-)}$) of both $C_3N$ and graphene are mainly determined by optical phonon branches, as can be seen from Fig. 5(e-f). The biggest difference between the emission processes of $C_3N$ and graphene is that the optical phonon branches of $C_3N$ show much more emission channels than graphene. So it is certain that the emission process O→LA/TA/LA+O will play a critical role in the phonon transport of $C_3N$.

**3.4.2 Phonon anharmonicity**

The anharmonic nature of a certain structure can be roughly quantified by the Grüneisen parameter ($\gamma$).[36,40,49] In order to evaluate the phonon anharmonicities of monolayer $C_3N$ and graphene, we calculated the Grüneisen parameter ($\gamma$) of the two materials, as plotted in Fig 6(a-b). It is found that the Grüneisen parameters of the ZA branches show fully negative $\gamma$ for both materials, while the TA, LA and OP branches show both negative and partial positive $\gamma$. The large negative of ZA branches shares the general feature of 2D materials due to the membrane effect[50] but it should be noted that the scattering of ZA is largely suppressed due to the symmetry-based selection rule:[43,46] for one-atom-thick materials, reflection symmetry makes the third-order-force constants involving an odd number of out-of-plane direction vanish, as a result, scattering modes like ZA + ZA↔ZA, ZA + LA/TA↔LA/TA could never happen.

More importantly, the magnitude of $\gamma$ on $C_3N$ is slightly larger compared with that of graphene, especially for the three acoustic phonon branches, indicating stronger phonon anharmonicity in $C_3N$ sheet. The stronger phonon-phonon scattering due to the anharmonicity leads to the smaller phonon lifetime of $C_3N$ compared to graphene, as can be seen from Fig. 4(c-d), and thus leads to the lower $\kappa$.

In order to find more physical insight of the phonon anharmonicity in $C_3N$ and graphene, we further perform analysis from the view of electronic structures since all the properties are fundamentally determined by the atomic structure and the behavior of electrons. In Fig. 6(c-d), we plotted the electron localization function (ELF) of $C_3N$ and graphene. The higher electron localization on a particular bond consequently illustrates higher rigidity of that bonds[51] and such that the comparison of electron localization can provide useful information for the evaluation of the anharmonicity of $C_3N$ and graphene. It is seen that the high electron localizations of graphene occurring at the center of carbon-carbon bonds [Fig. 6(d)]. As for $C_3N$, the high electron localizations occurring at the center of carbon-carbon and carbon-nitrogen bonds, indicating the character of covalent bonds where the electrons are shared between two connecting atoms. Most importantly, the C-C and C-N covalent bonds can be easily identified from the ELF of $C_3N$. Additionally, we investigate the charge transfer from the Bader's charge analysis[52] and find that each C atom transfers about $1.31e$ to N atom in $C_3N$. It means that there is a significant polarizability of the C-N bonds. Naturally, there is no charge transfer among atoms for graphene because the bond is formed between the same atom types. Compared with other calculated charge tranfer of other canon nitrides such as penta-$CN_2$ ($1.2e$)[51] and nitrogen chain encapsulated in carbon nanotube ($0.4e$),[54] the N atoms in $C_3N$ receive much more charges, indicating a much polarizability of the C-N bonds and naturally stronger anharmonicity of this structure.

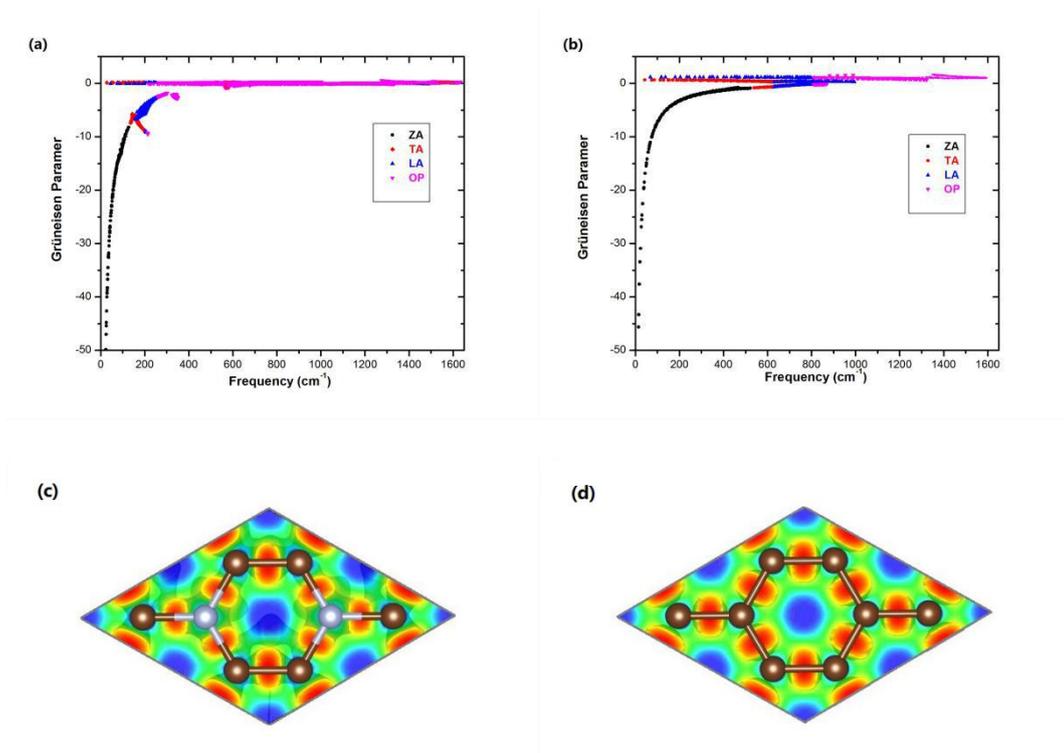

Fig. 6 Mode-Grüneisen parameters of (a) $C_3N$ and (b) graphene, and the electron localisation function (ELF) of (c) $C_3N$ and (d) graphene.

### 3.5 Size effect

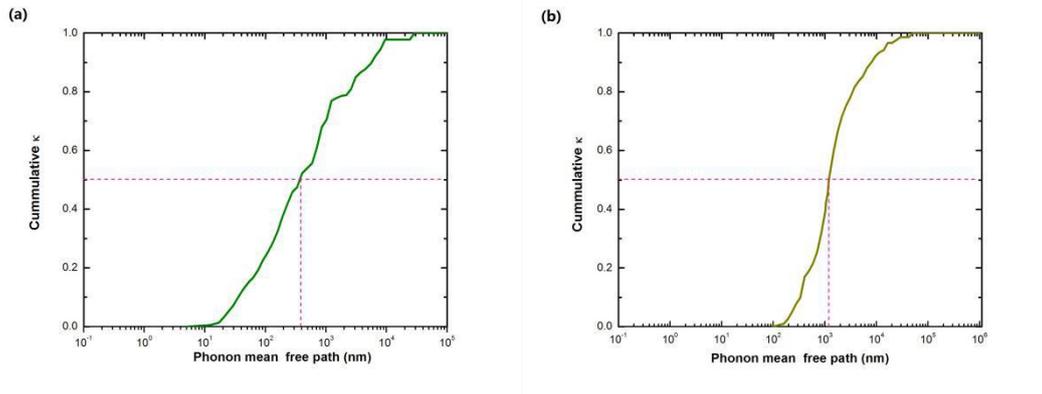

Fig. 7 Cumulative lattice thermal conductivity of (a) $C_3N$ and (b) graphene as a function of phonon mean free path (MFP) at 300 K.

Finally, we conduct some discussions on the phonon MFP and further the size effect on the phonon transport in the two materials. The accumulative $\kappa$ values with respect to MFP of $C_3N$ and graphene are plotted in Fig. 7(a-b). We can see that compared with graphene, $C_3N$ possesses a much broader phonon MFP spectrum ranging from a few nanometers to near $10^5$ nm. Furthermore, the contribution from the phonons with a short MFP is an important part of the total thermal transport for $C_3N$, which is quite different from graphene where the phonons with a long MFP contributed most to the thermal conductivity. By examining the 50% accumulated $\kappa$, the rMFP can be obtained. Our calculated rMPF of graphene at room temperature is 1250 nm, agrees well with the prior work (1164 nm at 300 K).[36] The rMPF of $C_3N$ is found to be 395 nm at 300 K, much lower than that of graphene, but is greatly larger than many typical 2D materials, for example, the rMPF of *h*-BN at 300 K is about 100 nm[36], and that of $C_2N$-*h*2D is only 36 nm.[28] It is obvious that when the length decreases the $\kappa$ of $C_3N$ will decrease faster with a larger rMFP. Additionally, with limited size, $\kappa$ of $C_3N$ could be effectively lowered by nanostructuring, such as patterning into nanoribbons or incorporating pores, which may extend their applications to thermoelectrics and thermal management.

## Conclusion

In summary, we conducted a comprehensive investigation of phonon transport property and lattice thermal conductivity of a recently experimentally grown single layer 2D material ($C_3N$) by using first-principles calculations coupled with phonon Boltzmann transport equation, and performed a comparison with graphene. Our DFT result reveals that the lattice thermal conductivity of $C_3N$ is as high as 380 W/mK, although much lower than that of graphene, the value is greatly higher than many other typical 2D materials. The three-phonon process in $C_3N$ is further analyzed and the results mean that the scattering between the acoustic and optical phonon modes like ZA/TA/LA+O→O processes plays a crucial role in the phonon transport of $C_3N$, which is quite different from graphene where the scattering channel of A+A→A is the primary three-phonon process in the thermal transport. The phonon anharmonicity of $C_3N$ is found much stronger than graphene and the result are further confirmed by the analysis from the view of electronic structures. At last, the size effect of $C_3N$ are compared with graphene, and found $C_3N$ possesses a much broader phonon MFP spectrum and much lower rMPF compared to graphene. Our prediction of the lattice thermal conductivity of $C_3N$ and the further analysis of underlying mechanisms indicate that semiconducting and high thermal conductivity of $C_3N$ sheets may have many potential in the field of thermal management concerns in various applications.

## Acknowledgements

This study was supported by the Natural Science Foundation of China (Grant no. 11547030, 11504155, 11704195 and 21363019).